\documentclass{SCIS2025}
\usepackage{epstopdf}

\begin{document}
\ArticleType{REVIEW}
\Year{2025}
\Month{}
\Vol{}
\No{}
\DOI{}
\ArtNo{}
\ReceiveDate{}
\ReviseDate{}
\AcceptDate{}
\OnlineDate{}
\AuthorMark{}
\AuthorCitation{}

\title{Advances in Continuous Variable Measurement-Device-Independent Quantum Key Distribution}{Title for citation}

\author[1]{Pu WANG}{}
\author[2]{Yan TIAN}{tianyan@nuc.edu.cn}
\author[3,4,5]{Yongmin LI}{yongmin@sxu.edu.cn}


\address[1]{School of Information, Shanxi University of Finance and Economics, Taiyuan 030006, China}
\address[2]{School of Information and Communication Engineering, North University of China, Taiyuan 030051, China}
\address[3]{State Key Laboratory of Quantum Optics and Quantum Optics Devices, Institute of Opto-Electronics,\\ Shanxi University, Taiyuan 030006, China}
\address[4]{Collaborative Innovation Center of Extreme Optics, Shanxi University, Taiyuan 030006, China}
\address[5]{Hefei National Laboratory, Hefei 230088, China}

\abstract{Continuous variable quantum key distribution (CV-QKD), utilizes continuous variables encoding such as the quadrature components of the quantized electromagnetic field and coherent detection decoding, offering good compatibility with the existing telecommunications technology and components. Continuous variable measurement-device-independent QKD (CV-MDI-QKD) can eliminate all the security threats arising from the receiver effectively, the crucial security loophole of CV-QKD implementations. Recently, CV-MDI-QKD has attracted extensive attentions and witnessed rapid progress. Here, we review the achievements that have been made in the field of CV-MDI-QKD, including the basic principle, advancements in theoretical protocols and experimental demonstrations. Finally, we discuss the challenges faced in practical applications and future research directions.}


\maketitle

\section{Introduction}
Quantum key distribution (QKD) is the most mature technology in quantum information processing, enabling two distant parties, Alice and Bob, to establish a common secret key over an insecure quantum channel with the aid of an authenticated classical channel~\cite{1,2,3,4}. The security of QKD is fundamentally guaranteed by the principles of quantum mechanics, ensuring that any eavesdropping attempts by Eve introduces detectable perturbations on the quantum states that carrying the key information~\cite{5,6,7,8}. Among various QKD protocols~\cite{9,10,11}, continuous variable (CV) QKD has garnered great research interests recently~\cite{12,13,14,15,16,17,18,19}, given its compatibility with the modern coherent optical communication techniques and its potential to achieve high secret key rates at metropolitan distances using multiphoton quantum states encoding and coherent detection (homodyne or heterodyne).

As a promising solution for secure communication in metropolitan quantum networks, CV-QKD systems have achieved notable achievements, with numerous protocols proposed and experimentally demonstrated over the past two decades~\cite{20,21,22,23,24,25,26,27,28,29,30,31,32,33,34,35,36,37,38,39,40,41,42,43,44,45,46,47,48,49,50,51,
52,53,54,55,56,57,58,59,60,61,62,63,64,65,66,67,68,69,70,71,72,73,74,75,76,77,78,79,80,81,82,83,84,85,86,87,88,
89,90,91,92,93,94,95,96,97,98,99}. Despite these achievements, practical implementation challenges still exist, especially the practical security of CV-QKD. Theoretically, the CV-QKD protocols have been proven to be information-theoretically secure under ideal conditions. However, real-world physical devices often deviate from these ideal assumptions, leading to potential security loopholes that can be exploited by adversaries. An effective solution is the device-independent (DI) QKD protocol~\cite{100,101}, aiming to eliminate all assumptions about the internal working mechanisms of QKD devices. However, it currently remains impractical due to low secret key rates and short transmission distances.

A more feasible solution came with the introduction of measurement-device-independent QKD (MDI-QKD)~\cite{102,103}, which removes all side-channel attacks on measurement devices, the most vulnerable part of QKD implementations. Moreover, it is particularly suitable for star-type metropolitan QKD networks. The concept of MDI-QKD was lately extended to the CV framework, so called CV-MDI-QKD~\cite{104,105,106}. In this protocol, Alice and Bob independently prepare CV quantum states and send them to an untrusted third party, Charlie, who performs CV Bell-state measurement (BSM) and broadcasts the outcomes. This protocol allows for the establishment of a secure key between Alice and Bob without relying on trusted detectors, thus closing known and unknown side-channel attacks on the detection side and significantly enhancing the practical security.

The inherent advantages of CV-MDI-QKD have attracted intense research attentions and witnessed rapid progress in recent years. This review is devoted to provide a comprehensive overview of the state-of-the-art in CV-MDI-QKD. We first delineate the procedures and fundamental principle of the CV-MDI-QKD protocol. Subsequently, we conduct an exhaustive review of the theoretical advancements in the field, encompassing protocol design and optimization, as well as in-depth security analysis. Furthermore, we present the recent proof-of-principle experimental validations. Finally, the review outlines the challenges and future research directions that lie ahead in the field, providing insights into the potential pathways for further advancements and developments.

\section{CV-MDI-QKD protocol}
\subsection{Protocol description}
\begin{figure}[htbp!]
\centering
\includegraphics[width=14cm]{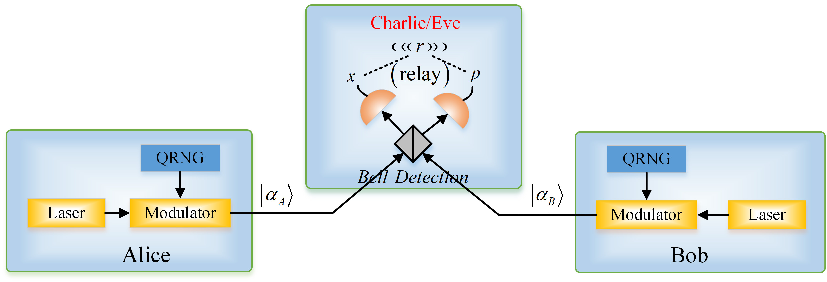}
\caption{(Color online) The PM scheme of the CV-MDI-QKD protocol with Gaussian-modulated coherent states.}
\label{fig1}
\end{figure}

When describing QKD protocols, two schemes are typically employed: "prepare-and-measure" (PM) and "entanglement-based" (EB). The PM scheme is usually easy to implement in practice, while the equivalent EB scheme is convenient for security analysis of the protocol. To understand how the CV-MDI-QKD protocol works, we start with the PM scheme involving Gaussian-modulated coherent states. It is one of the most widely used CV-MDI-QKD protocols and has been experimentally demonstrated~\cite{107}. The schematic setup is shown in Figure~\ref{fig1} and the protocol can be implemented by the following steps:

(1) At the transmitter, Alice and Bob, each independently encode the key information on the amplitude and phase quadratures of a series of coherent states $\left| {{\alpha _A}} \right\rangle $ and $\left| {{\alpha _B}} \right\rangle $ by using amplitude and phase modulators. In the phase space, the encoded states are expressed as $\left| {{\alpha _A}} \right\rangle  = \left| {{x_A} + i{p_A}} \right\rangle $ and $\left| {{\alpha _B}} \right\rangle  = \left| {{x_B} + i{p_B}} \right\rangle $, where ${x_A}$ and ${p_A}$ (${x_B}$ and ${p_B}$) represent two independent field quadratures with zero mean and identical variance ${V_A}$ (${V_B}$) in shot-noise units (SNUs). Subsequently, both Alice and Bob send their coherent states to an untrusted quantum relay, Charlie, via two unsecure lossy and noisy quantum channels.

(2) At the receiver, a CV BSM is performed. To this end, Charlie applies a beam splitter (BS) with a transmittance of 50\% to interfere the received signal states and establish the correlation. The output states are subsequently detected by using two homodyne detectors: one detects the amplitude quadrature and the other detects the phase quadrature, and the final measurement results are publicly declared by Charlie.

(3) Since Alice and Bob independently prepare their coherent states, whose complex amplitudes follow independent and identically distributed, zero-mean Gaussian distributions, their initial data sets are uncorrelated. To obtain a secret key, Alice and Bob apply a displacement operation to their data based on Charlie's measurement outcomes. Specifically, upon receiving Charlie's measurement results, Alice and Bob adjust their data as follows: ${X_A} = {x_A} - {g_{{x_A}}}\left( r \right)$, ${P_A} = {p_A} - {g_{{p_A}}}\left( r \right)$, ${X_B} = {x_B} - {g_{{x_B}}}\left( r \right)$, ${P_B} = {p_B} - {g_{{p_B}}}\left( r \right)$, where ${g_ * }$ ($ *  = {x_A},{p_A},{x_B},{p_B}$) represent the displacement coefficients that relate to Charlie's measurement results~\cite{107,108}. By conditionally displacing their data, Alice and Bob can achieve correlated data sets.

(4) Finally, by implementing parameter estimation, information reconciliation, and privacy amplification procedures, the secret keys can be extracted.

\subsection{Security analysis}
\begin{figure}[htbp!]
\centering
\includegraphics[width=14cm]{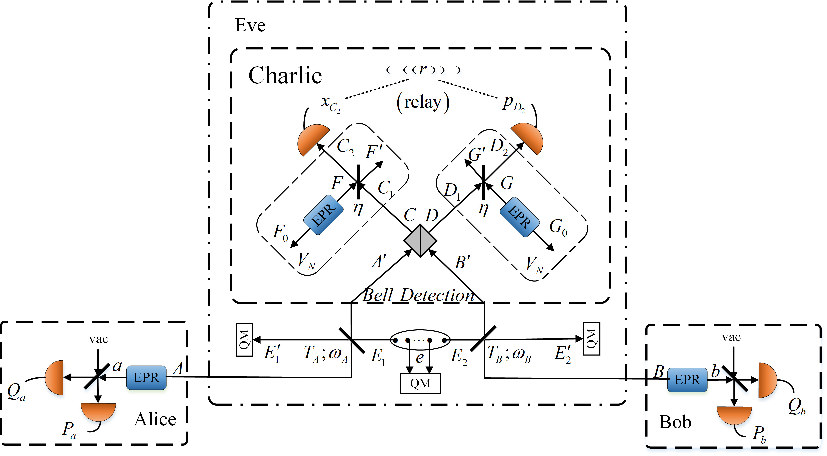}
\caption{(Color online) Equivalent EB scheme of the CV-MDI-QKD protocol with Gaussian-modulated coherent states.}
\label{fig2}
\end{figure}

The security of CV-MDI-QKD protocol can be established through the implementation of an equivalent EB scheme, as depicted in Figure~\ref{fig2}. In this scheme, instead of distributing the coherent states, Alice and Bob each generate an Einstein-Podolsky-Rosen (EPR) pair $aA$ or $bB$, respectively. Subsequently, they perform heterodyne detection on the retained mode $a$ or $b$, which projects mode $A$ or $B$ onto coherent states. Modes $A$ and $B$ are transmitted to a trusted third party, Charlie, via separate quantum channels with length ${L_{AC}}$ and ${L_{BC}}$, respectively. The received Modes $A'$ and $B'$ of Charlie interfere at a BS and output two modes $C$ and $D$. Then the $x$ quadrature of mode $C$ and the $p$ quadrature of mode $D$ are measured using balanced homodyne detectors, respectively. The realistic homodyne detectors are modeled by assuming that the signal is attenuated by a BS with transmission efficiency $\eta $ and mixed with some thermal noise ${V_N}$ which simulates the electronic noise ${v_{el}}$ of the detector, before detected by a perfect homodyne detector. Charlie publicly announces the complex variable $r=\left ( x_{C_{2}}+ip_{D_{2} } \right )/ \sqrt{2}$  to both Alice and Bob through an authenticated classical channel. Here, the knowledge of $r$  enables them to infer each other's data through the previously discussed data processing techniques. Consequently, a correlation between Alice and Bob is established and results in mutual information ${I_{ab\left| r \right.}} > 0$. Finally, the secret keys are extracted via classical data post-processing techniques including the parameter estimation, information reconciliation, and privacy amplification.

It is worth noting that a realistic joint two-mode Gaussian attack can be performed by Eve on the two quantum channels of the CV-MDI-QKD. As illustrated in Figure~\ref{fig2}, Eve mixes two ancillary modes, denoted as ${E_1}$ and ${E_2}$, with two incoming modes, $A$ and $B$, respectively, through two BSs with a transmittance of ${T_A}$ and ${T_B}$. A thorough analysis of the two-mode attacks reveals that, in the asymptotic limit, the most effective attack is the "negative EPR attack"~\cite{109,110,106,111}.

Since the protocol is symmetric, for convenience, we assume that Alice is the encoder and Bob is the decoder. After declaration of Charlie's outcome $r$, the asymptotical secret key rate (i.e., for raw keys of infinite length) against collective attacks is given by~\cite{106}
\begin{equation}
 {K^\infty } = \beta  \cdot {I_{ab\left| r \right.}} - {\chi _{aE\left| r \right.}},
\end{equation}
where $\beta $  denotes the reconciliation efficiency; ${I_{ab\left| r \right.}}$ denotes the Shannon mutual information between Alice and Bob; and ${\chi _{aE\left| r \right.}}$ denotes the Holevo bound between Alice and Eve, which puts an upper limit on the information available to Eve on Alice's key. Based on the purification of Eve's system, ${\chi _{aE\left| r \right.}}$ can be calculated by the von Neumann entropy of the quantum states ${\rho _{ab\left| r \right.}}$ and ${\rho _{b\left| {ra} \right.}}$.

In practice, there are two application scenarios for the protocol: the symmetric configuration (${L_{AC}} = {L_{BC}}$) and the asymmetric configuration (${L_{AC}} \ne {L_{BC}}$). In the symmetric configuration, the maximum achievable distance is about 3.8 km of standard optical fiber from the relay with present technology~\cite{106}. The performance of the asymmetric case is superior to the symmetric case. When Alice is the encoder of information, the transmission distance ${L_{BC}}$ increases significantly as ${L_{AC}}$ decreases. For the most asymmetric case (${L_{AC}} = 0$), a key rate of $2\times10^{-4}$ bit/pulse can be achieved at a distance of 170 km under ideal conditions~\cite{112}.

By including the finite data statistics effect for parameter estimation and the post-processing costs, the security of CV-MDI-QKD protocol with Gaussian-modulated coherent states is extended to the finite-size scenario under realistic conditions~\cite{113,114}. Recently, by involving smooth min-entropy and Gaussian de Finetti reduction, the composable security of the protocol against coherent attacks is established~\cite{115,116}.

\section{Theoretical advances of CV-MDI-QKD}
After the first Gaussian-modulated coherent states CV-MDI-QKD protocol was proposed, many efforts have been made to improve the protocol. For instance, various protocols have been developed to reduce the complexity of the system, enhance the performance of the protocol, and ensure the realistic security.

The unidimensional modulation, discrete modulation and passive-state preparation schemes were introduced to reduce the complexity of CV-MDI-QKD system. In the unidimensional modulation scheme~\cite{117}, both Alice and Bob use one modulator to implement the single-quadrature modulation, then the prepared quantum states are sent to Charlie for BSM. In this case, the signal modulations as well as the corresponding parameter estimations can be simplified. The discrete modulation~\cite{118,119}, for example, four-state scheme, further simplifies the state preparation and allows a good reconciliation efficiency at low signal-to-noise ratio. Apart from the active state preparation, passive-state preparation is also an attractive alternative for the practical application of CV-MDI-QKD protocol~\cite{120,121}, where both Alice and Bob passively prepare quantum states using a true thermal source.

To enhance the performance of the original protocol in terms of the secret key rate and distance, a variety of schemes have been proposed. The squeezed states scheme was introduced into the CV-MDI framework to attain better performance~\cite{122,123}. Later, the modulated squeezed states~\cite{124} that combining the advantages of both the squeezed and coherent states was proposed, it can achieve a higher secret key rate and transmission distance than previous protocols. Besides, other schemes including multi-mode modulation~\cite{125}, one-time shot-noise-unit calibration~\cite{126}, optical amplifier~\cite{127,124}, photon subtraction~\cite{128,129,130,131}, quantum catalysis~\cite{132,133,134}, quantum scissor~\cite{135}, and postselection~\cite{136}, have also been studied.

A key security assumption in MDI-QKD is that the source is trusted. Even though one can prepare the source with good fidelity in practice, there are inevitably some preparation errors. There are several works that use different approaches to prove the realistic security of the imperfect state preparation~\cite{137,138,139}. Recently, a countermeasure for negative impact introduced by the actual source in the CV-MDI-QKD system based on the one-time-calibration method was proposed~\cite{140}. To solve the Local Oscillator (LO) transmission, the plug-and-play (P\&P) technique~\cite{141,142} and Bayesian phase-noise estimation technique~\cite{143} were introduced, eliminating the need for transmitting high-intensity LOs. Additionally, researchers also analyzed the performance of CV-MDI-QKD protocols under various complex communication environments, such as fluctuating channel transmittance~\cite{144}, rainy and foggy weather environment~\cite{145}, underwater communication~\cite{146}, satellite-to-submarine model~\cite{147}, and free-space optical links~\cite{148}. These results provide useful guidance for practical applications.

\begin{figure}[htbp!]
\centering
\includegraphics[width=14.5cm]{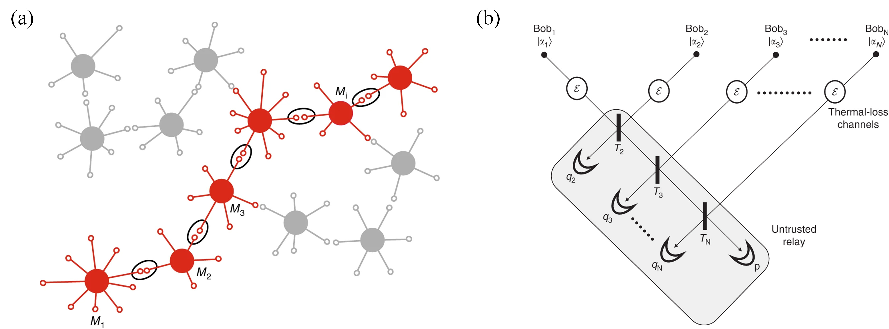}
\caption{(Color online) (a) Modular network for secure quantum conferencing. (b)  Each module $M_i$ is a star network comprise of a central untrusted relay and $N_i$ trusted users, which is a generalization of CV-MDI-QKD~\cite{149}. Copyright 2019 Springer Nature.}
\label{fig3}
\end{figure}

Another interesting application of CV-MDI-QKD is quantum conferencing~\cite{149,150}, where multiple parties can securely share information in a group setting. Figure~\ref{fig3}(a) plots the modular network model for quantum conferencing, where each module $M_i$ represents a star network as shown in Figure~\ref{fig3}(b)~\cite{149}. Two different modules can be connected by a pair of trusted users. In each star-network module, each of $N_i$ users prepares their own signal states according to the Gaussian distribution and send them to an untrusted relay node via quantum channels, where a multipartite CV Bell detection is performed. After the measurement outcomes are broadcasted, all users in module $M_i$ reconcile their data with a trusted user that is shared with another module $M_j$. As the distance from the central relay increases or the number of users rises, the secret key rate for each star network decreases. In ideal conditions, within a radius of 40 meters, a typical distance for a large building, 50 users can communicate privately with a secret key rate exceeding 0.1 bit per signal use.

\section{Experimental advances of CV-MDI-QKD}
\begin{figure}[htbp!]
\centering
\includegraphics[width=12cm]{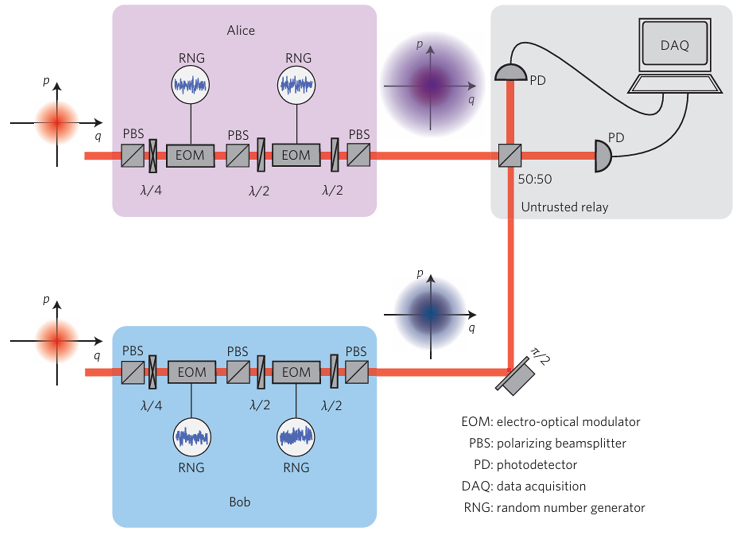}
\caption{(Color online) The proof-of-principle demonstration of CV-MDI-QKD~\cite{106}. Copyright 2015 Springer Nature.}
\label{fig4}
\end{figure}

In 2015, the first proof-of-principle demonstration of CV-MDI-QKD was reported~\cite{106}, as shown in Figure~\ref{fig4}. A high-stability continuous wave laser at 1064 nm was divided into two parts and delivered to Alice and Bob, respectively. Both Alice and Bob use amplitude and phase modulators to modulate the light field independently with zero centered Gaussian distributions in phase space. Subsequently, the signal fields were transmitted to the receiver's site, Charlie, through free space. At Charlie's site, the signal fields sent by Alice and Bob interfere at a free-space 50:50 BS, and a pair of conjugate quadratures of the output fields are measured by two high-efficiency balanced homodyne detectors. The quantum signal was encoded on the sidemode, and the carrier of the laser beam was used as the LO. Thus, the continuous-variable Bell-state measurement was significantly simplified by directly subtracting and adding the measurement results of the balanced BS output modes, which produce the difference of the amplitude quadratures and the sum of the phase quadratures. The losses in the links were simulated by varying the variances of the modulation signals. With a reconciliation efficiency of 97\% and a total quantum efficiency of 98\%, the secret key rate achieved in this experiment is three orders of magnitude higher than that of the qubit-based protocols over metropolitan area, providing a promising solution of building high-rate metropolitan quantum networks.

The crucial issue that makes the long distance CV-MDI-QKD challenging is the implementation of high-efficiency CV-BSM of two remote independent quantum states, which required the establishment of a reliable phase reference between two spatially separated lasers. The dual-homodyne detection is required to achieve simultaneous measurement of a pair of conjugate quadratures. Besides, the imperfect detection efficiency at Charlie's site is equivalent to optical losses that inevitably induce vacuum fluctuation noises, which, along with detector electronic noises, both contribute to the untrusted noises. Hence, the performance of CV-MDI-QKD heavily depends on the detection efficiency of Charlie's detectors, which requires high efficiency photodiodes and low transmission loss.

\begin{figure}[htbp!]
\centering
\includegraphics[width=14cm]{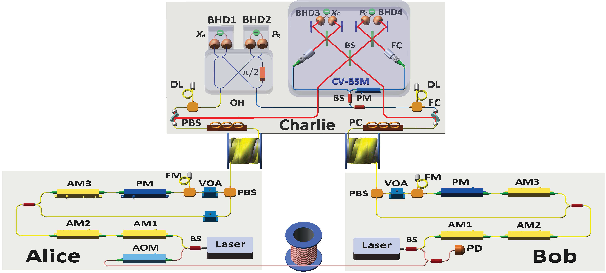}
\caption{(Color online) Experimental demonstration of continuous-variable measurement-device-independent quantum key distribution over long distance optical fiber~\cite{107}. Copyright 2022 Optica Publishing Group.}
\label{fig5}
\end{figure}

In 2022, the experimental demonstration of CV-MDI-QKD over long distance optical fiber was realized~\cite{107}, where a technology that consists of optical phase locking, phase estimation, real-time phase feedback, and quadrature remapping was developed to accurately implement CV-BSM of remote independent quantum states, as shown in Figure~\ref{fig5}. Two single-frequency continuous-wave lasers with linewidth of kHz at 1550 nm were employed by Alice and Bob. An optical phase-locked loop technique was adopted to compensate the frequency difference of the two independent lasers, where part of Alice's laser beam is frequency-up-shifted by 80MHz and sent to Bob's station, which interferes with part of Bob's laser beam to generate a beat signal for frequency-locking. Subsequently, Both Alice and Bob adopt two cascaded amplitude modulators to generate 50-ns light pulses with a repetition rate of 500 kHz, and modulate the signal pulses independently and randomly with zero-centered Gaussian distributions in phase space. In order to accurately estimate the slow phase drifts of the signal and phase-reference (LO) fields in real time, Alice and Bob periodically insert some phase-calibration pulses into the signal pulses. Finally, the signal and phase-reference (LO) fields are time and polarization multiplexed, and sent to Charlie through SMF-28 fiber spools.

At receiver's site, a 90-deg optical hybrid was used to perform heterodyne detection and obtain the amplitude and phase quadrature of the phase-reference pulses to estimate the fast phase shift for each signal pulse. The fast phase shift consists of residual phase noise after frequency-locking of two independent lasers, the phase noise arising from the finite laser linewidth, and independent transmission fiber links. A phase modulator was adopted to apply a compensated phase shift to one of the LO fields in CV-BSM based on the values estimated by the phase-calibration pulses of Alice and Bob in real time, to ensure that the field quadratures measured by dual-homodyne detection were faithfully orthogonal. Besides, to achieve high-efficiency CV-BSM, the signal and LO fields are coupled from optical fibers to free space where high quality free-space optical components with very low losses are used. Two time-domain BHDs with a quantum efficiency of 99\% are developed to measure two conjugated quadratures. Considering the insertion loss of the optical components and interference visibility, the total detection efficiency is 97.2\%. Finally, both Alice and Bob implement a quadrature remapping, where they rotate their data at hand with the estimated phase-drift information to ensure that the data measured by Charlie to be matched with the data of Alice and Bob. With a reconciliation efficiency of 97\%, the distance between Bob (Alice) and Charlie of 0.1 km (5/10 km), the achieved secret key rate is 0.43 (0.19) bits/pulse. When the transmission distance is less than 15 km, the secret key rate of the CV protocol is significantly better than that of its DV counterpart even considering the cryogenic single-photon detectors. The proposed approaches in this work comprise a promising solution for construction of a high key rate and low-cost metropolitan CV-MDI-QKD network.

\begin{figure}[htbp!]
\centering
\includegraphics[width=14cm]{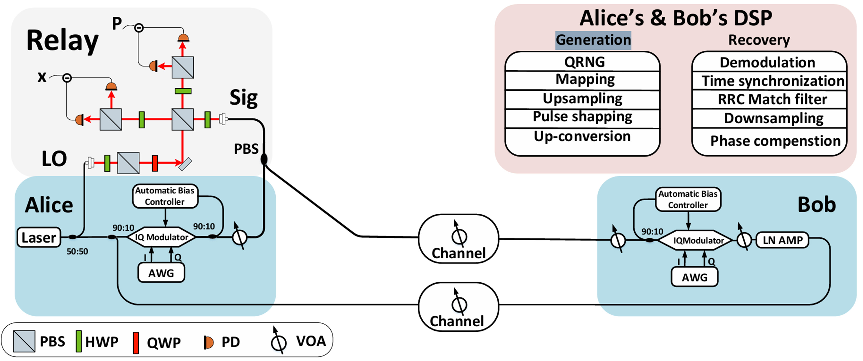}
\caption{(Color online) CV-MDI-QKD system without frequency and phase locking~\cite{151}. Copyright 2023 Optica Publishing Group.}
\label{fig6}
\end{figure}

In 2023, a simple and practical CV-MDI-QKD system was reported, which was achieved by using a new relay structure leveraging the concept of a polarization-based 90-degree optical hybrid and digital signal processing (DSP) pipeline for CV-BSM~\cite{151}, as shown in Figure~\ref{fig6}. A 1550 nm continuous-wave laser at Alice with a linewidth of 100 Hz was shared with the relay to implement an asymmetric configuration of the CV-MDI protocol, where the relay and Alice were placed together. Moreover, a portion of Alice's laser was sent to Bob through an independent fiber channel in order to avoid the frequency locking.

In each transmitter, an in-phase and quadrature (IQ) modulator was used to generate the ensemble of coherent states. The DSP techniques consisting of digital pulse shaping and sideband modulation were implemented, simplifying the CV-MDI-QKD system. At the relay, the incoming signal beams from Alice and Bob were overlapped at a fiber-based polarization beamsplitter. Then, the signal was coupled into the free-space polarization-based 90-degree hybrid. Because the LO was prepared in circular polarization by using a quarter wave-plate and the signal was linearly polarized, the amplitude quadrature and phase quadrature can be detected simultaneously. After the relay publicly announces the output of CV-BSM, the quantum signals were recovered using DSP. The relay output was digitally demodulated to baseband by downconversion and low-pass filtering. Temporal synchronization was achieved by calculating the cross-correlation between the samples transmitted by Alice and Bob and the relay output. Then, the synchronized samples were matched filtered and downsampled to symbols. In order to compensate for the phase drift, both Alice and Bob rotated their data at hand to maximize the cross-correlation. Finally, Alice and Bob performed displacement operations on their own data according to the relay output to correlate their data. Considering the information reconciliation efficiency of 97\%, total detection efficiency of 94\%, and Bob's channel loss of 2 dB, a secret key fraction of 0.12 bit per relay use can be achieved.

\begin{figure}[!t]
\centering
\includegraphics[width=14cm]{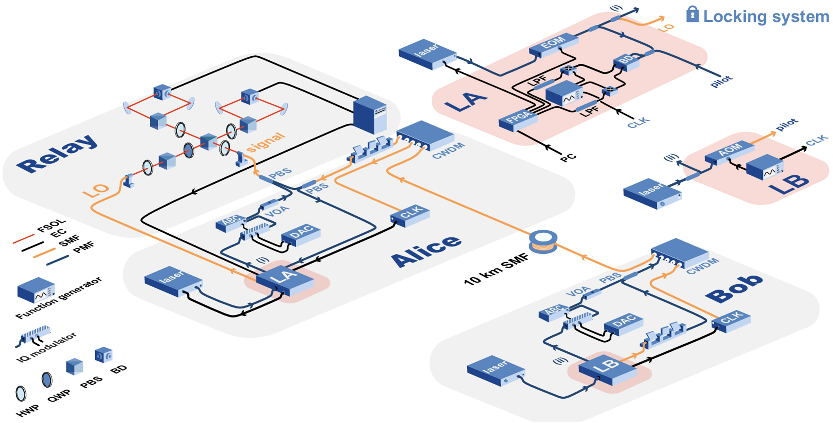}
\caption{(Color online) Experimental CV-MDI-QKD with finite-size security security against collective attacks~\cite{152}. Copyright 2023 arXiv.}
\label{fig7}
\end{figure}

In 2024, Hajomer et al. further performed the experimental demonstration of CV-MDI-QKD with finite-size security against collective attacks. In this work, a locally generated LO based on a real-time phase locking system were adopted~\cite{152} in comparison to their previous work, as shown in Figure~\ref{fig7}. An asymmetric configuration of the MDI protocol was implemented, where the relay is co-located with Alice's station. Bob's station and the relay were connected through a single-mode fiber.

To experimentally realize CV-BSM over a long distance, a heterodyne optical locking system was employed to phase-lock the two independent lasers that are used to generate the quantum states at Alice's and Bob's stations. At Bob's station, a part of the laser beam was frequency shifted by 40 MHz using an acousto-optical modulator and then sent to Alice's station, where it was interfered with part of Alice's laser beam on a 50:50 BS. The beat signal generated by the interference was detected by a balanced detector. The phase detection was performed by analog I-Q demodulation at 40 MHz. An FPGA was used to generate an error signal, and drive the piezoelectric wavelength modulator inside Alice's laser to compensate for the slow phase fluctuations. An electro-optic phase modulator was adopted to compensate for the fast phase fluctuations. Finally, Alice's stabilized laser was used as the optical source of quantum signal and shared with the relay as LO through a short single-mode fiber channel.

Alice's and Bob's stations were clock synchronized using a synchronizing clock signal. To this end, a 10 MHz master clock generated at Bob's station was converted to an optical signal at a wavelength of 1310 nm using an electrical-to-optical converter circuit. Subsequently, the optical clock was multiplexed with the quantum signal using a Coarse Wavelength Division Multiplexer (CWDM) and transmitted to Alice's station through the same fiber channel. At Alice's station, the optical clock was then converted back to an electrical signal and distributed to the DAC, the locking system and the relay's ADCs. Considering the information reconciliation efficiency of 97\%, total detection efficiency of 94\%, a block size of $4\times 10^{6}$, and a failure probability of $10^{-10}$, a positive expected secrete key rate of 2.6 Mbit/s was achieved over 10 km fiber link.

The common phase-reference is the crucial challenge for CV-MDI-QKD because of the CV-BSM of two remote independent quantum states. By placing the laser at Charlie's site and using plug-and-play configuration~\cite{141}, the issues of synchronization between different lasers as well as the generation of LO can be solved. Moreover, the polarization drifts can be compensated automatically since only one laser is needed. Yin et al. proposed a phase self-aligned CV-MDI-QKD scheme~\cite{153}. By delicately manipulating the polarization state of the quantum signals, they can transmit along the same fiber link before CV-BSM in the relay. Thus their relative phase fluctuation can be negligible and the phase-reference is self-aligned. This approach enables that the reliable phase reference can be established.

Above schemes effectively solve the phase-reference problem, however, there are still some issues that need to be addressed. The first is the untrusted source problem, which exists in all plug-and-play type QKD systems. For example, the Trojan-horse attack will greatly decrease the key rate along with the increasing of the mean photon number of the Trojan-horse mode. The other one is noise photons caused by the strong Rayleigh scattering, which will affect the coherent detection at Charlie.

\section{Challenges and Future Directions}
Despite significant progress, practical CV-MDI-QKD still faces technical challenges. Present protocols in general require very high detection efficiency for the heterodyne detection. Furthermore, the best performance only achieves at an asymmetric configuration and the distribution distance is still limited for a (nearly) symmetric one. Although a variety of schemes have been proposed to improve the performance of CV-MDI-QKD, the experimental verification is still required. Future research needs to focus on designing new protocols that can operate using heterodyne detection with ordinary detection efficiency. Furthermore, they should show superior performance in both asymmetric and symmetric configurations.

With the rapid progress of photonics integrated technology, the on-chip integrated CV-MDI-QKD system is crucial to meet the demands of miniaturization, low power consumption, and low-cost ~\cite{154,155,156,157,158,159,160,161,162,163,164}, which are prerequisites for furture large-scale applications. In addition, in order to find more applications in quantum networks, it is desired to extend the CV-MDI-QKD protocols to multi-party in terms of specific application scenarios~\cite{165,166,167,168,169,170,171,172,173,174}.

\section{Conclusion}
In summary, this review provides a comprehensive overview of the past advancements in the field of CV-MDI-QKD. At present, both the asymptotic and composable security of the protocol have been rigorously proven. The experimental demonstrations over long distance optical fiber paves the way for the practical deployment of CV-MDI-QKD in real-world scenarios, particularly in metropolitan areas. Although significant progresses have been made, challenges still remain for future practical applications. Both the theory and technique breakthroughs are crucial to overcome the obstacles and fully realize the promise of CV-MDI-QKD in practical applications.

\Acknowledgements{This work was supported by the National Natural Science Foundation of China (Grant Nos. 62175138, 62205188 and 62305198), Shanxi 1331KSC, Innovation Program for Quantum Science and Technology (Grant No. 2021ZD0300703), Fundamental Research Program of Shanxi Province (Grant Nos. 20210302124290, 202303021212168 and 202403021212343), and Scientific and Technologial Innovation Programs of Higher Education Institutions in Shanxi (STIP) (Grant No. 2024L183).}




\end{document}